\documentclass[preprint,aps,graphicx]{revtex4}
\usepackage{graphicx}
\usepackage{dcolumn}
\usepackage{bm}

\begin{document}

\title{Collisions of bosonic ultracold polar molecules in  microwave traps.}

\author{Alexander V. Avdeenkov$^{1,2}$}
\affiliation{1.
National Institute for Theoretical Physics (NITheP),
Stellenbosch Institute of Advanced Study, Stellenbosch 7600, South Africa
\\
2. Institute of Nuclear Physics, Moscow State
University, Vorob'evy Gory, Moscow, 119992 Russia}


\begin{abstract}
The collisions between linear polar molecules, trapped in a
microwave field with circular polarization, are theoretically
analyzed. The microwave trap suggested by DeMille~\cite{DeMille}
seems to be rather advantageous in comparison with other traps. Here
we have demonstrated that the microwave trap can provide a
successful evaporative cooling for polar molecules in a rather broad range of
frequencies of the AC-field.
We suggested that not only ground state polar molecules but also molecules in some other states
can be safely trapped.
 But the state in which molecules can be safely loaded and
trapped depends on the frequency of the AC-field.
\end{abstract}

\maketitle

\section{Introduction}
In the last few years a considerable progress has been made in the
development and investigation of various types of electromagnetic
traps for different but especially polar molecules. The building of
such traps should have made it possible to perform many interesting
experimental studies and applications involving the manipulation of
cooled and trapped molecules with kinetic temperatures near zero
Kelvin~\cite{zoller1,xia,zoller2,zoller3,hudson,lev,micheli,avd1}.
 Schemes for creating samples of cold molecules are being developed
now in several laboratories around the world  where simple
molecules at kinetic temperatures in the $mK$ to $\mu K$ range are
being created and studied. One of the main purposes in the
perfection of traps is to keep samples of molecules dense and cold
enough and the microwave trap seems to be very advantageous in
restraining polar molecules~\cite{DeMille}.

In this article we are studying the collisional dynamics of linear
polar molecules in the $^1\Sigma$ state trapped in a microwave field
with a circular polarization at cold temperatures. The choice of linear polarization
does not seem  very appealing as it was shown in~\cite{DeMille} as AC- Stark levels have a lot of
avoiding crossings which enhances the collisional loss.
Polar molecules have strong permanent electric dipole moments and therefore the collisional
dynamics is mostly ruled by the dipole-dipole interaction:
\begin{eqnarray}
V_{\mu \mu}(\vec{R})=\frac{\vec{\mu}_1\vec{\mu}_2-3(\vec{e}_R\vec{\mu}_1)(\vec{e}_R\vec{\mu}_2) }{R^3}
\end{eqnarray}
Here, $\vec{\mu}_i$ is the electric dipole moment of the molecule $i$, $\vec{R}$ is the intermolecular separation and
$ \vec{e}_R=\vec{R}/R$.
Thus in our model collisions are controlled by two ratios $\nu /B$ and $x=\mu {\cal E}/hB$~($\nu $ is the microwave frequency, $B$ is the
molecular rotational constant, and
${\cal E}$ is the electric field strength). We are using the
coupled-channels method to calculate elastic and inelastic cross
sections in dependence on the frequency and the strength of an AC
electric field. It has to be noted that collisional dynamics of cold polar molecules
in the DC- electric field was rather intensively theoretically explored which revealed quite rich
 physics~\cite{ticknor1, avd2,bohn,avd3,avd4,ticknor2,doyle,krems}

 One of the main goals
is to understand if polar molecules in their absolute ground state
can have a successful evaporative cooling. At first sight it seems
rather an inappropriate question as molecules are in their absolute
ground state. Naturally, if the energy of the first excited
rotational state is smaller than the microwave frequency($\nu <B$)
it seems that the system is safe against two-body inelastic
collisions and collision losses are excluded. And this is true for
bare particles. But it might be not completely true for particles
dressed with a microwave field , especially for parameters which
provide a large and deep trap($\nu \sim B,\mu {\cal E} \gtrsim B$).
As we have seen the reason is that the AC- field and the
dipole-dipole interaction are mixing states with different angular
moments and "dressing". We have found that even for the most
favourable case $\nu <B$, there are inelastic processes. But these
processes are causing mostly  the "undressing" or "dressing" of the colliding
particles without changing their internal structure. Here we have to
be careful about the term "mostly" as dressed states are the superposition of lots of pure states.
But  what we can be sure of is that these "dressing"-"undressing" processes do not cause the loss
of molecules from the trap.

 Moreover this is
supposedly be true not only for molecules in their absolute ground
state $|J=0,M=0>$, but for example in the $|J=1,M=-1>$ state at
frequencies around $\nu/B=3$, for molecules $|J=2, M=-2>$ at
frequencies around $\nu/B=5$ and so on($J$-rotational quantum
number). Actually how far this "around" goes depends on the strength
of the AC- field.

We have found that during the collisions the parity of the
projection of the total momentum of two colliding dressed  molecules
is conserved. Our calculations demonstrate that the cross
sections for such positive and negative projection parities are
different by several orders of magnitude.

\section{Polar $^1 \Sigma$- type molecules in a microwave field}

Here we analyze the AC-Stark shift, which determines the trap
depth.
 The energy levels of $^{1}\Sigma $- type molecules can be
described by
rotation $J$,  total spin $F$ (i.e., including nuclear spin), and vibration $\upsilon $ 
quantum numbers. For simplicity we will neglect
hyperfine splitting and consider  molecules only in the $\upsilon
=0$ vibrational ground state. So we  treat polar molecules as rigid
rotors with a permanent dipole moment. The dressed- state formalism is particularly
convenient for describing the photon-atom interaction in the strong field limit~\cite{coh,wei} where
the dressed states are eigenstates of the Hamiltonian of the total system: particle plus photon.
The basis states of the rigid rotor plus field Hamiltonian, $H=H_{rot}+H_{field}$ are $\left| J,M, \bar{N}+n\right\rangle $, where $M$ is the
projection of $J$ on space-fixed axis, which is conveniently  chosen parallel to the wave-vector of the microwave field.  , $\bar{N}+n$ is the dressed state photon number and $n<<\bar{N}$ is the deviation of the photon
number. The AC- Stark splitting is caused by the molecule- field interaction $H_{Stark}=-\vec{\mu}\vec{\cal E}$.
Considering the circularly polarized microwave field ($\sigma
^{-}$), the non-zero Hamiltonian matrix elements (normalized by
$\hbar B$) are given by
\begin{eqnarray}
\left\langle J,M,n\left| H_{rot}+H_{field} \right| J,M,n\right\rangle  =J\left(
J+1\right)
+n\frac{\nu }{B}   \\
\nonumber \left\langle J+1,M+1,n+1\left| H_{Stark} \right| J,M,n\right\rangle
=
\frac{x}{2}\frac{\sqrt{J+M+1}\sqrt{J+M+2}}{2\sqrt{2J+1}\sqrt{2J+3}}
\\
\left\langle J+1,M-1,n-1\left| H_{Stark} \right| J,M,n\right\rangle
=
\frac{x}{2}\frac{\sqrt{J-M+1}\sqrt{J-M+2}}{2\sqrt{2J+1}\sqrt{2J+3}},
\end{eqnarray}
Here, it is assumed that $n$ is much smaller than the mean photon
number $N$. In the AC- electric field the $J, M, N+n$ are not good
quantum numbers and the dressed state formalism should be applied.

\subsection{Dressed states}
\begin{figure}
\centerline{\includegraphics[width=0.9\linewidth,height=0.7
\linewidth,angle=0]{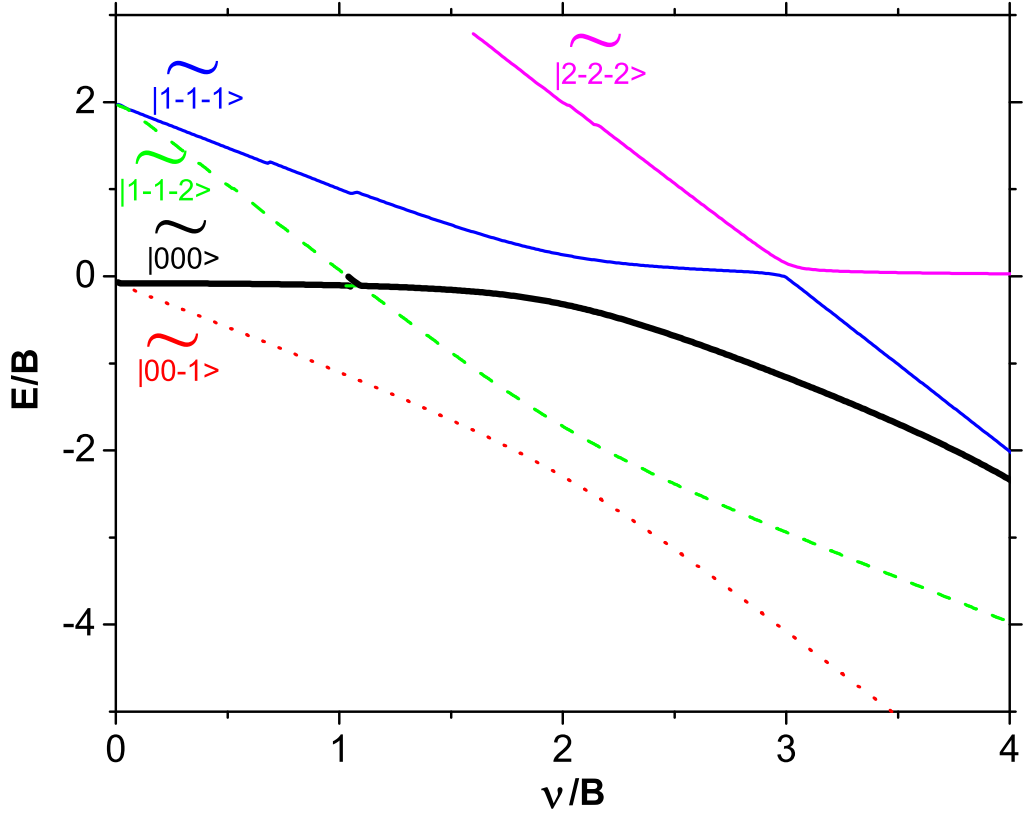}} \caption{ Energies of dressed states
versus applied microwave electric field frequency at $x=0.7$. See
text for details. } \label{stark}
\end{figure}

The AC- electric field mixes states with different $J$, $M$ and $n$ and hence neither of them is a good quantum number.
Thus we can only mark our states by their origin at zero field(Fig.\ref{stark}) where they can be assigned by $J$ and $M$ quantum numbers.
Figure \ref{stark} shows the energies of dressed states versus
applied electric field frequency for the $J=0$ and $J=1$ states  at
$x=0.7$ within  $\nu /B=[0,4]$ frequency range.

First we would like to consider the ground and so the lowest- field seeking state $|J=0,M=0>$.
Though  the term lowest-field seeking state  does not seem appropriate in dressed state formalism.
In practice we transform the
molecular state to a field-dressed basis for performing scattering
calculations and the state with a given mean photon
number $\bar{N}$  is described
as:
\begin{eqnarray}
\left|\widetilde{(JMn)}\right\rangle \equiv
\left|\widetilde{(JMn)};\vec{\cal E}\right\rangle  =\sum_{JMn} p_{\left( JMn\right)} \left|
JMn\right\rangle ,
\label{mix}
\end{eqnarray}
where $p_{\left( JMn\right)}$ stands for the eigenfunctions of the $H_{rot}+H_{field}+H_{Stark}$ Hamiltonian
determined numerically at each value of the field.We will
continue to refer to the molecular states by the quantum
numbers $J$, $M$, and $n$, with the understanding that they are
only approximately good in a field, and that Eq.~\ref{mix} is the
appropriate molecular state.
Note that AC- Stark splitting should be irrelevant to the value of $|n|$ as $n<<\bar{N}$ so we choose small values of $n$.
Further we are mostly interested in cases in which the strength of the AC- field is not large($\mu {\cal E} \lesssim B$) which is found to be more practical for a microwave trap~\cite{DeMille}. In this case the above mentioned mixing is not that strong and allows us to analyze different cases.

\subsection{$\nu \lesssim  2B$ case}
Let us consider  $\left|\widetilde{(000)}\right\rangle$ state which is a pure $\left|000 \right\rangle $ if we neglect the molecule- field interaction $H_{Stark}$ and  describe molecules in their absolute ground state.
In~\cite{DeMille} it was pointed out that the most relevant trap parameters are at $x\gtrsim 1$ and the ratio
$\nu/B \sim 1$. In this case the state of interest is approximately described as:
\begin{eqnarray}
\left|\widetilde{(000)}\right\rangle  \approx p_{\left( 000\right)} \left| 000\right\rangle +
p_{\left( 1-1-1\right)} \left| 1-1-1\right\rangle
\label{mix000}
\end{eqnarray}
It has to be noted that at $x\gtrsim 1$ ,  $p_{\left( 000\right)} \approx p_{\left( 1-1-1\right)} \approx 1/\sqrt{2}$.
At larger $x$ admixtures of other states, like $\left| 2-2-2\right\rangle$ and so on are not small.
Here we would like to point out that if  trapped molecules are in this state $\left|\widetilde{(000)}\right\rangle$ it is equal to saying that trapped molecules are in
$\left|\widetilde{(00|n|)}\right\rangle$ states as the AC Stark shift is the same for all of them.
Moreover the $H_{rot}+H_{field}+H_{Stark}$ Hamiltonian does not mix, for example, states like  $\left|\widetilde{(000)}\right\rangle$ and $\left|\widetilde{(00\pm 1)}\right\rangle$.
But the dipole- dipole interaction can mix them  as the  $$\left|\widetilde{(000)}\right\rangle \left|\widetilde{(000)}\right\rangle \Rightarrow  \left|\widetilde{(00\pm 1)}\right\rangle \left|\widetilde{(00\pm 1)}\right\rangle  $$ scattering processes are possible.
In order to reveal the possibility of such processes one has to see that at  $x\gtrsim 1$ and the ratio
$\nu/B \sim 1$ these states are mostly described as:
\begin{eqnarray}
\left|\widetilde{(00-1)}\right\rangle  \approx p_{\left( 00-1\right)} \left| 00-1\right\rangle +
p_{\left( 1-1-2\right)} \left| 1-1-2\right\rangle \\
\left|\widetilde{(001)}\right\rangle  \approx p_{\left( 001\right)} \left| 001\right\rangle +
p_{\left( 1-10\right)} \left| 1-10\right\rangle
\label{mix001}
\end{eqnarray}
Thus the
\begin{eqnarray}
\left\langle \widetilde{(000)}\right|\mu \left|\widetilde{(00 -1)}\right\rangle \sim  p_{\left( 1-1-1\right)} p_{\left( 00-1\right)} \left\langle 1-1-1\right|\mu \left| 00-1\right\rangle
\\
\nonumber
\left\langle \widetilde{(000)}\right|\mu \left|\widetilde{(00 1)}\right\rangle \sim  p_{\left( 000\right)} p_{\left( 1-10\right)} \left\langle 000\right|\mu \left| 1-10\right\rangle
\end{eqnarray}
matrix elements provide for the above-mentioned process and their values depend on $p_{(JMn)}$ eigenstates.

At $\nu \lesssim B$ the only two inelastic processes may have considerable cross sections:
\begin{eqnarray}
\left|\widetilde{(000)}\right\rangle \left|\widetilde{(000)}\right\rangle \Rightarrow  \left|\widetilde{(00- 1)}\right\rangle \left|\widetilde{(001)}\right\rangle
\\
\nonumber
\left|\widetilde{(000)}\right\rangle \left|\widetilde{(000)}\right\rangle \Rightarrow  \left|\widetilde{(00- 1)}\right\rangle \left|\widetilde{(00-1)}\right\rangle
\end{eqnarray}
Note that the threshold energies of $\left|\widetilde{(000)}\right\rangle \left|\widetilde{(000)}\right\rangle$ and $\left|\widetilde{(00- 1)}\right\rangle \left|\widetilde{(001)}\right\rangle$ channels are degenerate.

At $\nu \gtrsim B$ the $\left|\widetilde{(1-1-2)}\right\rangle $ state  becomes energetically accessible(see Fig.~\ref{stark}) so that the
 \begin{eqnarray}
 \left|\widetilde{(000)}\right\rangle \left|\widetilde{(000)}\right\rangle \Rightarrow  \left|\widetilde{(00- 1)}\right\rangle \left|\widetilde{(1-1-2)}\right\rangle
 \end{eqnarray}
 inelastic process is possible.

\subsection{$2B \lesssim \nu \lesssim 4B$ case}
In this region of frequencies our state of interest $\left|\widetilde{(000)}\right\rangle$
is approximately described as:
 \begin{eqnarray}
\left|\widetilde{(000)}\right\rangle  \approx p_{\left( 000\right)} \left| 000\right\rangle +
p_{\left( 1-1-1\right)} \left| 1-1-1\right\rangle + p_{\left( 2-2-2\right)} \left| 2-2-2\right\rangle
\label{mix002}
\end{eqnarray}
 If the strength of the field is not large(see Fig.~\ref{stark}) then in the middle of this region this state is an almost pure $\left| 1-1-1\right\rangle$ state.
 As $\nu /B$
gets close to five,  the state is dominated by the
$\left|2,-2,-2\right\rangle $ state. So the larger the frequency of
the AC- electric field, the larger particular angular moment $J$ will
be dominated. But this is only true  if the strength of a field is
not large($x\lesssim 1$). Otherwise, at a large $x$, the dressed state
will be a superposition of a lot of states.

 Compared to the $\nu \lesssim  2B$ case there is one extra open channel for inelastic collisions:
 \begin{eqnarray}
 \left|\widetilde{(000)}\right\rangle \left|\widetilde{(000)}\right\rangle \Rightarrow  \left|\widetilde{(1-1- 2)}\right\rangle \left|\widetilde{(1-1-2)}\right\rangle .
 \end{eqnarray}
 Moreover our state in this region of frequencies becomes a weak-field seeking one while the state
 $\left|\widetilde{(1-1-2)}\right\rangle$ becomes a strong-field seeking state.

\section{Results and discussions}
The main difference from
the DC electric field case is that in general  inelastic processes
are always allowed   for any non-zero field and for any state. Of course an inelastic
cross section will be very small at small fields but not exactly
zero. In the following we focus only on bosonic species. The case of fermionic
species will be considered elsewhere.

We have solved the coupled- channel equations using the logarithmic
derivative propagator method \cite{Johnson} to calculate total
state-to-state cross sections.The scattering calculations quickly
become computationally expensive as more rotational states, partial
waves and dressed-state photon numbers are included. We have found
that it is enough to take $J=0,1$; $l=0,2$ for bosons and $l=1,3$ for
fermions and $\Delta n =0,\pm1,2$ for changing  photon numbers to
get reasonably converged cross sections. The only two-body
interaction taken into account here is the dipole-dipole interaction
which is sufficient to describe the qualitative behavior of cross
sections.

We have not chosen any particular polar molecule for calculation but
a rigid rotor with the rotational constant $B=6GHz$ and dipole
moment $\mu=3D$. For this molecule, we are mostly interested in the
lowest energy strong-field-seeking state of the ground vibrational
state. The cross sections were calculated for dressed molecules,
which means we show them for molecules in
$\left|\widetilde{(JMn)};\right\rangle$ basis. The trap depth for
this particular molecule is $12mK$ at $x=0.5$ and $50mK$ at $x=1$ if
$\nu/B =1/3$. It is $60mK$ at $x=1$ and $\nu/B =1$. For the high
frequency case $\nu/B =2.5$ it is only $3.5mK$ at $x=1$.

We have found that if one takes into account only the dipole-dipole
interaction in the presence of an AC-field  the only conserved value
for two molecules in
$\left|\widetilde{(J_1M_1n_1)};\widetilde{(J_2M_2n_2)}\right\rangle$
states  is $(-1)^{M_1+M_2+m_l+n_1+n_2}$ which we call a projection
parity. Thus the system of the coupled- channel equations is split
into two blocks with a positive and negative projection parity. The
nature of this symmetry will be described elsewhere.

\begin{figure}
\centerline{\includegraphics[width=1.1\linewidth,height=0.8
\linewidth,angle=0]{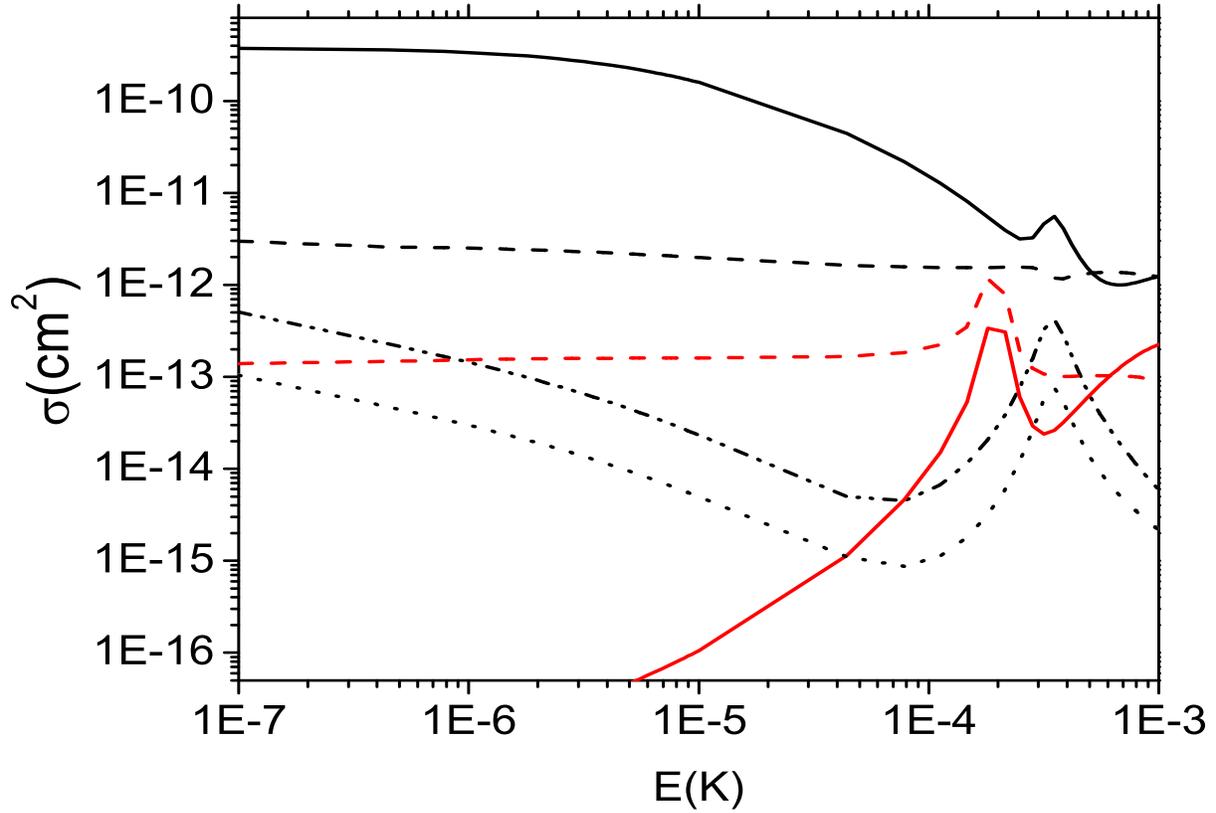}}
 \caption{Elastic(solid lines) and inelastic (not solid lines) cross
sections versus the collisional energy for $x\approx 1$, $\nu/B \approx 1$. The black and red curves are for the
positive and negative projection parities respectively. The dashed curves are for
$\left|\widetilde{(000)}\right\rangle
\left|\widetilde{(000)}\right\rangle \Rightarrow
\left|\widetilde{(00- 1)}\right\rangle
\left|\widetilde{(001)}\right\rangle$ ,
the dotted curve is for
$\left|\widetilde{(000)}\right\rangle
\left|\widetilde{(000)}\right\rangle \Rightarrow
\left|\widetilde{(00-1)}\right\rangle
\left|\widetilde{(00-1)}\right\rangle $
 and the dot-dot-dashed curve is for
$\left|\widetilde{(000)}\right\rangle
\left|\widetilde{(000)}\right\rangle \Rightarrow
\left|\widetilde{(00- 1)}\right\rangle
\left|\widetilde{(1-1-2)}\right\rangle $ transitions.
}
 \label{f6e4}
\end{figure}

Fig. \ref{f6e4} shows cross sections for bosonic and fermionic
species for the $\left|\widetilde{(000)};\right\rangle$ state at $x=1$
and $\nu/B=1$ parameters which are proposed as more relevant for
the future experiment~\cite{DeMille}. The black and red curves are for
positive and negative projection parities respectively. The largest
inelastic cross section is for $\left|\widetilde{(000)}\right\rangle
\left|\widetilde{(000)}\right\rangle \Rightarrow
\left|\widetilde{(00- 1)}\right\rangle
\left|\widetilde{(001)}\right\rangle$ transitions and moreover it is
not small and the threshold behavior is similar to that of an elastic process.
The latter is clear as these states are degenerate.
Such inelastic transition  can be described as the "dressing" or
"undressing" of molecules by one photon each. Will this "undressing" cause the loss of
molecules from the trap? We believe that it will not as in the limit
of a strong field the Stark shift as well as the trap depth will be
the same for dressed states with $\bar{N}$ and $\bar{N} \pm 1$ photons.
The only
inelastic process which causes the trap loss is
$\left|\widetilde{(000)}\right\rangle
\left|\widetilde{(000)}\right\rangle \Rightarrow
\left|\widetilde{(00- 1)}\right\rangle
\left|\widetilde{(1-1-2)}\right\rangle $. But the cross section for
it is relatively small in the region from about $0.1\mu K$ to $100\mu K$ and
even maybe at lower temperatures. One can see that there  exists some resonance structure.
We do not take any notice of  these resonances as they are defined by an unknown short-range part
of the interaction which in our case is simply modeled by the inner wall of the interaction.
Thus after all our "absolute ground state" molecules undergo inelastic transitions. Though at
$x\approx  1$ and $\nu/B \approx 1$ trapped molecules will feel  pretty safe.

 It is
noteworthy to note that the elastic cross sections are much larger
for positive parities  at
ultracold energies(around $1\mu K$) especially while they are more
or less of the same order at cold energies(around $1mK$). We have
found that the threshold behavior of the inelastic cross section for
bosonic molecules with a negative projection parity does not follow
the Wigner law at the shown collisional energies at least. The only
reason we can think of is that for a given projection parity not all
projections $m_l$ of the partial wave $l$ are involved. This
phenomenon will be studied elsewhere.

\begin{figure}
\centerline{\includegraphics[width=0.55\linewidth,height=0.39
\linewidth,angle=0]{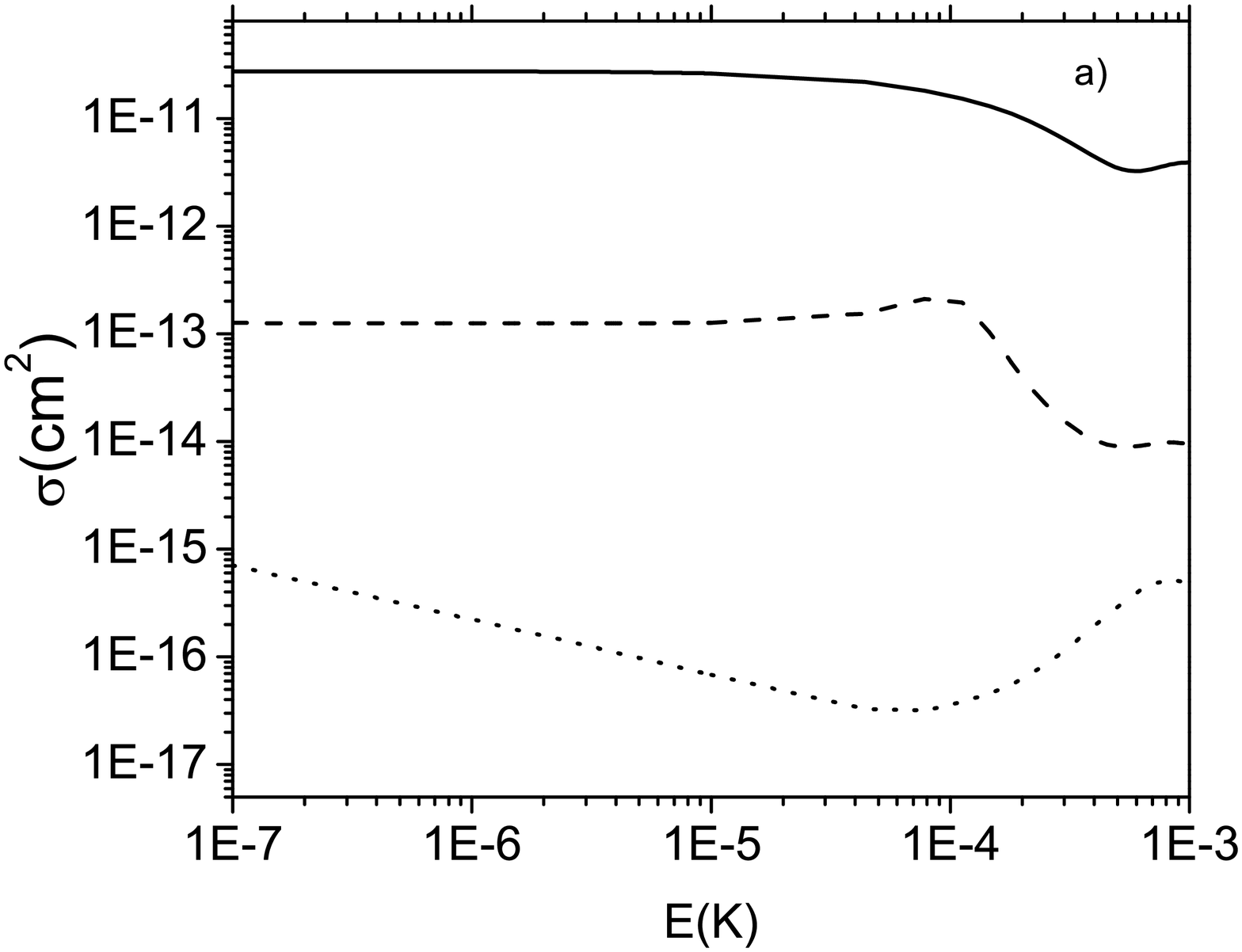}}
\centerline{\includegraphics[width=0.55\linewidth,height=0.39
\linewidth,angle=0]{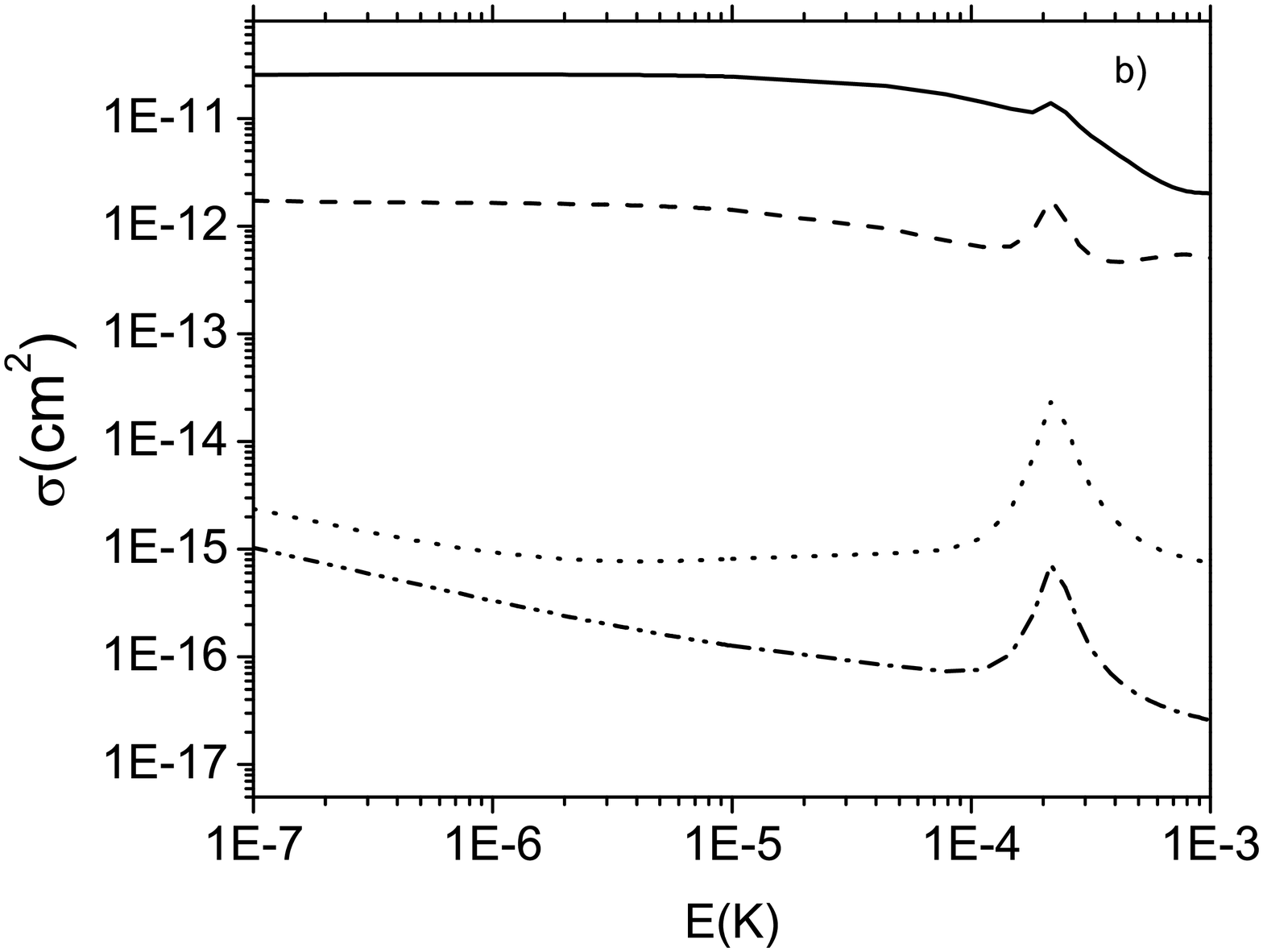}}
\centerline{\includegraphics[width=0.55\linewidth,height=0.39
\linewidth,angle=0]{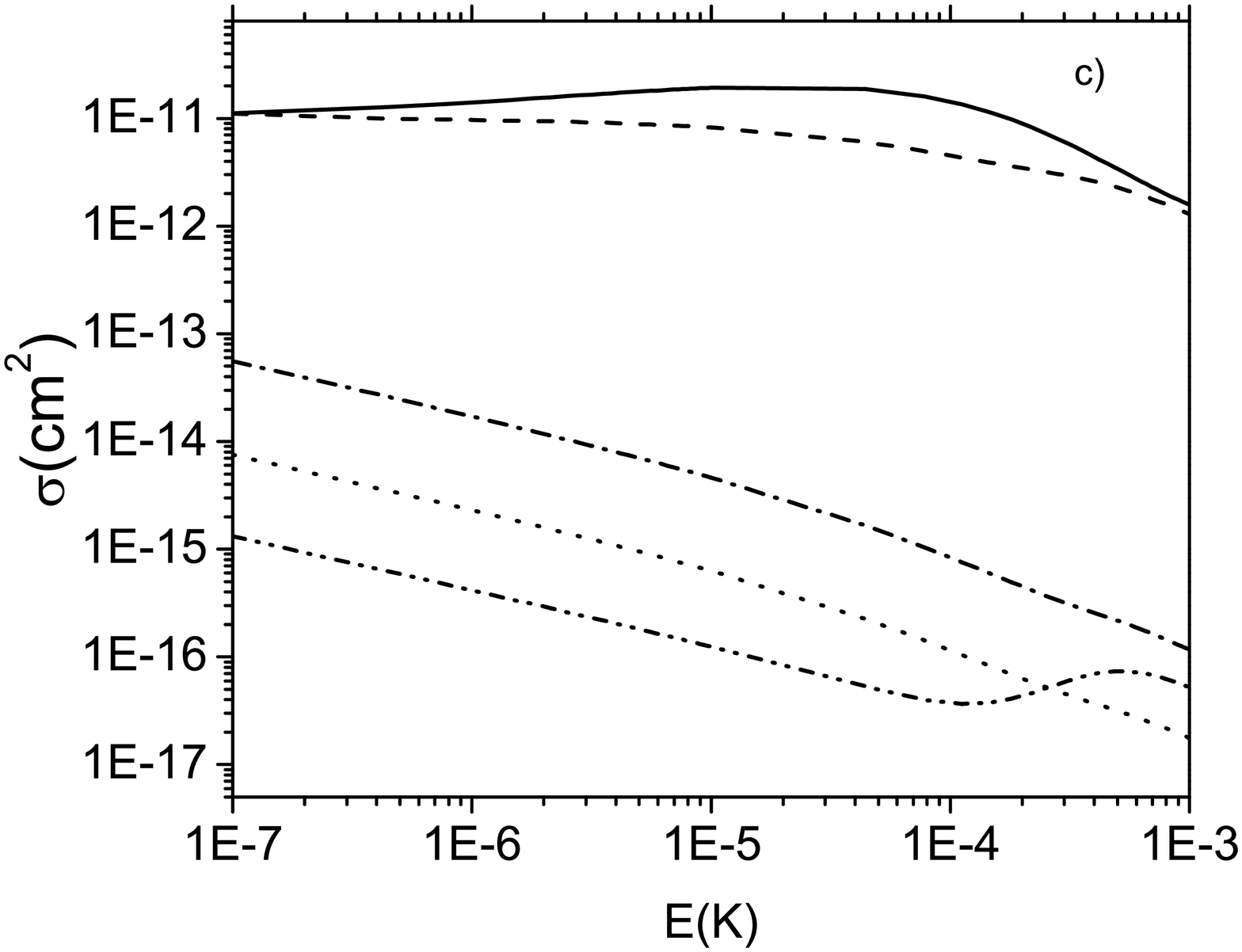}}
 \caption{ Elastic(solid lines) and inelastic (not solid lines) cross
sections versus the collisional energy at $x\approx 0.5$ for a)$\nu/B=0.3$, b) $1$
and c) $2.5$.
The dashed curves are for
$\left|\widetilde{(000)}\right\rangle
\left|\widetilde{(000)}\right\rangle \Rightarrow
\left|\widetilde{(00- 1)}\right\rangle
\left|\widetilde{(001)}\right\rangle$ ,
the dotted curves are for
$\left|\widetilde{(000)}\right\rangle
\left|\widetilde{(000)}\right\rangle \Rightarrow
\left|\widetilde{(00-1)}\right\rangle
\left|\widetilde{(00-1)}\right\rangle $,
 the dot-dot-dashed curves are for
$\left|\widetilde{(000)}\right\rangle
\left|\widetilde{(000)}\right\rangle \Rightarrow
\left|\widetilde{(00- 1)}\right\rangle
\left|\widetilde{(1-1-2)}\right\rangle $,
and the dot-dashed curve is for
$\left|\widetilde{(000)}\right\rangle
\left|\widetilde{(000)}\right\rangle \Rightarrow
\left|\widetilde{(1-1-2)}\right\rangle
\left|\widetilde{(1-1-2)}\right\rangle$
 transitions.}
 \label{e2}
\end{figure}

Fig. \ref{e2} shows cross sections for $x=0.5$ and
different $\nu/B$ ratios . Though the trap depth is not that deep for this $x$ it is still several $mK$.
We would like to pay attention to the relatively small value of the field strength as in this case  dressed states are mostly superpositions only of only two pure states or even sometimes one pure state~(see discussions in Section II).
In this case it is easy to analyze hidden details.
At $\nu/B \approx 0.3$ ratio the field is rather weak so  dressed states are almost pure states(
for example, it means that, say, $\left|\widetilde{(00-1)}\right\rangle \approx \left|(00-1)\right\rangle $,
see Eq.(5-8)) and all
inelastic cross sections are small.
At $\nu/B \approx 1$ ratio the dressed states are mostly superpositions of two states(Eq.(5-8)) which considerably enhance
"dressing"-"undressing" inelastic processes involving considerable transitions to $|J=1,M=-1>$ states by means of the
dipole-dipole interaction(Eq.8).  Nevertheless such transitions are between equally trapped states $\left|\widetilde{(00|n|)}\right\rangle$.
At $\nu/B=2.5$ ratio the dressed state of interest is by the largest part the $\left|(1-1-1)\right\rangle$ state and
it is already a weak-field seeking state.
Moreover one can see that inelastic cross sections which may cause the loss from a trap are really small.
We may suppose that it is possible to prepare polar molecules in their $|J=1,M=-1>$ states and then load them into
a microwave field at a frequency $\nu$ around $3B$. It is probable that the same trick can be done
with, say, $J=2,M=-2$ molecules by loading them into a microwave field at a frequency $\nu$ around $5B$.

\begin{figure}
\centerline{\includegraphics[width=1.1\linewidth,height=0.8
\linewidth,angle=0]{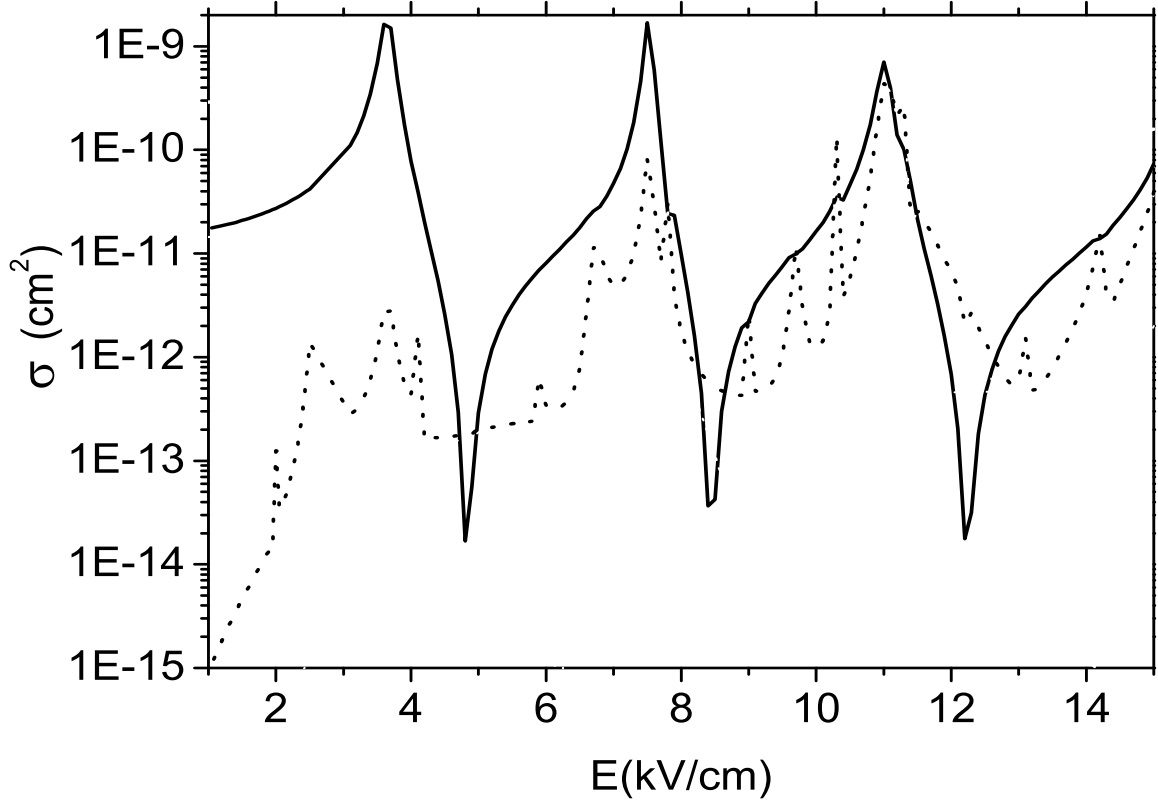}}
 \caption{Elastic(solid line) and total inelastic (dotted line) cross
sections
versus the microwave electric field strength at the collision energy $1\mu K$ for
$\nu/B=0.3$. The value $x \approx 1$ is attained at ${\cal E}=4kV/cm$. }
 \label{f2}
\end{figure}

\begin{figure}
\centerline{\includegraphics[width=1.1\linewidth,height=0.8
\linewidth,angle=0]{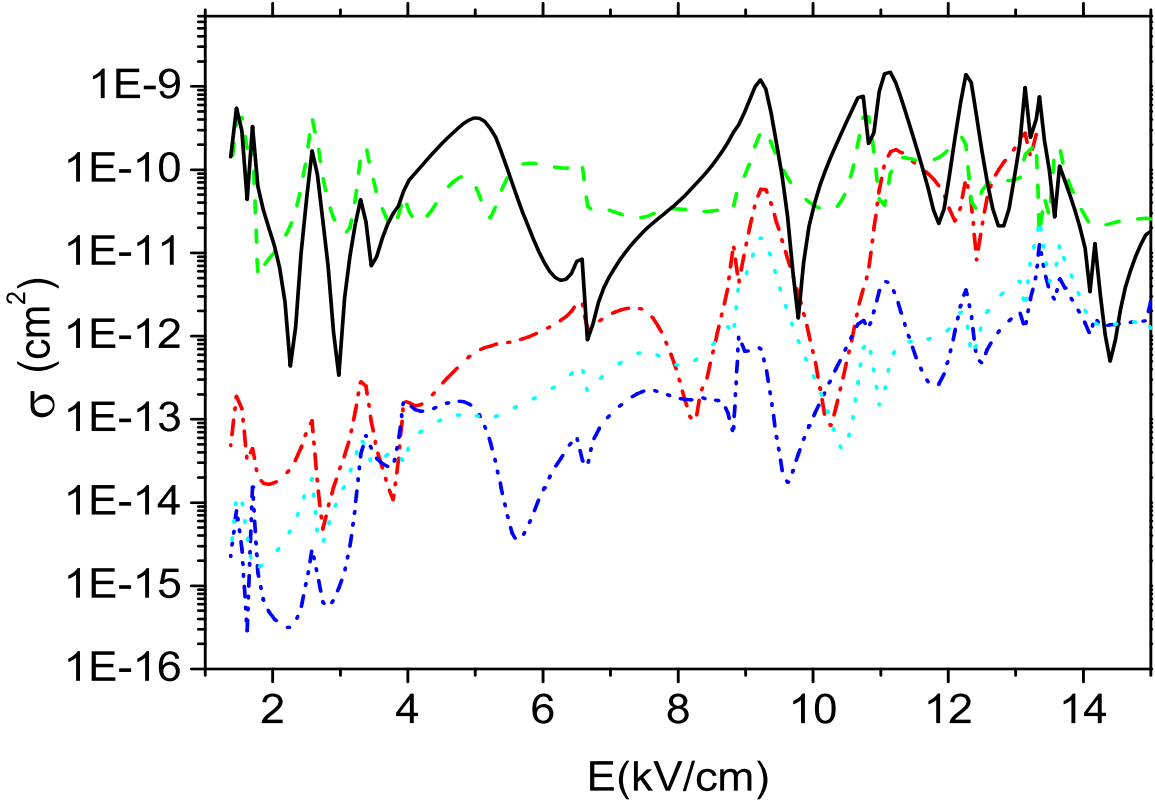}}
 \caption{Elastic(solid line) and inelastic (not solid lines) cross
sections versus
the microwave electric field strength at the collision energy $1\mu K$ for
$\nu/B=2.5$. The inelastic processes are marked as on Fig. \ref{e2}.}
 \label{f15}
\end{figure}

We have found out that the cross sections have rather a weak
dependence on the frequency of an AC-field(Fig.\ref{f2},\ref{f15}) for large values of the
strength of a field($x\approx 1$ at $E\approx 4kV/cm$).
Both
elastic and inelastic cross sections are more or less of the same
order. Of course there is rather a rich
resonance structure of cross sections which reflects the fact of the
existence of lots of bound states in closed channels.  We do not
pay attention to this fact as the short-range potential is unknown
and simply parameterized by the inner wall. But we would like to note that the positions of
resonances will depend not on the strength of a field(like it is for DC-field~\cite{ticknor})
but also on a frequency. Thus simply changing the frequency of a field one can catch a resonance.
This issue will be under our investigation elsewhere.
Fig.\ref{f2} shows that at large values of field the inelastic "dressing"-"undressing" cross sections become
comparable with  the elastic one but the low frequency($\nu/B=0.3$) does not allow other trap losses processes.
Thus the evaporative cooling may be successful for rather large strengths of AC-field as well. But it has to be
checked as for a larger field we have to take a larger Hilbert space of basis states $|JMn>$. It means
that some other inelastic channels with larger $J$,$M$ and $n$ may appear at a larger field and "destroy" such
 a pleasant picture. But it will be very costly numerically as we will have to handle up to several thousands of coupled-channels equations.
Fig.\ref{f15} shows the same pattern of behavior for a higher frequency. The only difference is that there already are
some inelastic channels at large fields which definitely cause trap losses.

To recapitulate, we have computed scattering cross sections for cold
collisions of the bosonic  polar $^1 \Sigma$- type
molecules as functions of both collision energy and AC electric
field parameters. We found that even ground state molecules have inelastic transitions.
Though most of them are harmless and allow keeping molecules in a trap.
We suggested that polar molecules at $|J,M=-J>$
states may have a successful evaporative cooling in the appropriate
region of frequencies of a microwave field and will be restrained
and shielded by this field against collision losses.

We acknowledge useful interactions with Masatoshi Kajita and Masato
Nakamura.


\section*{References}
\begin{thebibliography}{10}
\bibitem{DeMille} D. DeMille, D. R. Glenn, and J. Petricka; Euro. Phys. J.
D \textbf{31}, 375 (2004).
\bibitem{zoller1} Margareta Wallquist, Peter Rabl, Mikhail D Lukin
and Peter Zoller; New.J.Phys.\textbf{10}, 063005(2008).
\bibitem{xia}Yong Xia, Yaling Yin, Haibo Chen, Lianzhong Deng, and Jianping
Yin, Phys.Rev.Lett.\textbf{100}, 043003 (2008).
\bibitem{zoller2} A.Micheli, G.Pupillo, H.P.Buchler and
P.Zoller,Phys.Rev.A.\textbf{76}, 043604(2008).
\bibitem{zoller3} G. Pupillo,A. Griessner, A. Micheli, M. Ortner, D.-W. Wang and P.
Zoller, Phys.Rev.Lett.\textbf{100}, 050402 (2008).
\bibitem{hudson}E. R. Hudson et al., Phys. Rev. Lett. \textbf{96}, 143004(2006).
\bibitem{lev} B. L. Lev et al., Phys. Rev. A \textbf{74}, 061402(R) (2006).
\bibitem{micheli} A. Micheli et al., Nature Physics \textbf{2}, 341 (2006).
\bibitem{avd1} A. V. Avdeenkov, D. C. E. Bortolotti, and J. L. Bohn, Phys.
Rev.A. \textbf{69}, 012710 (2004).
\bibitem{ticknor1} C.Ticknor and J.L.Bohn, Phys.Rev.A.\textbf{72}, 022709 (2005).
\bibitem{avd2} A.V.Avdeenkov,M.Kajita and J.L.Bohn, Phys.Rev.A.\textbf{73}022707(2006).
\bibitem{bohn} J. L. Bohn, Phys. Rev.A \textbf{63}, 052714 (2001).
\bibitem{avd3} A. V. Avdeenkov and J. L. Bohn, Phys. Rev. A \textbf{66}, 052718 (2002).
\bibitem{avd4} A. V. Avdeenkov and J. L. Bohn, Phys. Rev. A \textbf{71}, 022706 (2005).
\bibitem{ticknor2} C.Ticknor, Phys. Rev. Lett. \textbf{100}, 133202(2008).
\bibitem{doyle} J. Doyle, B. Friedrich, R.V. Krems and F. Masnou-Seeuws, Eur.Phys.J. D, \textbf{31}, 149 (2004).
\bibitem{krems} R.V. Krems, Phys. Rev. Lett. \textbf{96}, 123202 (2006)
\bibitem{roud} V.Roudnev and M.Cavagnero, Phys. Rev. A \textbf{79}, 014701 (2009).
\bibitem{coh} C.Cohen-Tannoudji and S.Reynaud, J.Phys.B \textbf{10},345;365;2311(1977).
\bibitem{wei} Changjiang Wei et. al., Phys.Rev.A.\textbf{58}, 2310 (1988).
\bibitem{Johnson} B. R. Johnson, J. Comput. Phys. {\bf 13}, 445.
\bibitem{ticknor} C.Ticknor and J.L.Bohn, Phys.Rev.A.\textbf{72}, 032717 (2005).
(1973).
\end{thebibliography}
\end{document}